\documentclass{article}
\usepackage{amssymb}
\newcommand{\ud}{\mathrm{d}}

\title{Quantum Inverse Scattering Method and (Super)Conformal Field Theory}
\author{ Petr P. Kulish$^1$, Anton M. Zeitlin$^2$\\
$^1$ St. Petersburg Division\\
of Steklov Mathematical Institute,\\ Fontanka 27,
Saint-Petersburg, 191023, Russia\\
$^2$ Department of High Energy Physics, Physics Faculty,\\ 
St. Petersburg State University,
 \\ Ulyanovskaya 1, Petrodvoretz, St. Petersburg, 198904, 
Russia\\}
\date{}

\begin{document}
\maketitle
\begin{abstract}
In this paper we consider the possibility of application of the quantum
inverse scattering method for studying the superconformal field theory
and it's integrable perturbations.
The classical limit of the considered constructions is based on 
$\widehat{osp}(1|2)$ super-KdV hierarchy.
The quantum counterpart of the monodromy matrix  corresponding to the linear
problem associated with the L-operator
is introduced. Using the explicit form of the irreducible representations of
$\widehat{osp}_q(1|2)$, the ``fusion relations'' for the transfer-matrices 
(i.e. the traces of the monodromy matrices in different representations) are obtained.
\end{abstract}
\section{Introduction}
\subsection{Quantum Inverse Scattering Method}
Quantum Inverse Scattering Method (QISM) appeared in the works of the 
Leningrad school of mathematical physics in the late 1970-s \cite {skl79},\cite {kul79}. It has arisen as a synthesis of two approaches to the integrable systems. The first approach, the so-called Inverse Scattering Method (ISM) was discovered in 1967  \cite {GGKM} but has deep roots in the works on classical mechanics in 19-th century and the second approach was applied to the problems of statistical physics on a lattice and 
quantum mechanics \cite {Bethe},\cite {Baxter} up to the end of 1970-s. ISM gave possibility to find
classes (hierarchies) of the integrable two-dimensional nonlinear evolution equations and get their solutions. In a few years after ISM was discovered, 
the algebraic structure of this method was understood 
\cite {Lax} and the Hamiltonian interpretation was obtained \cite {zahfadd}.
 It appeared that the Hamiltonian systems, corresponding to these equations
are fully integrable and have infinite number of conservation laws.
 The algebraic structure of ISM allows to consider the integrable 
nonlinear equation under study as a compatibility condition of the 
system of linear equations:
\begin{eqnarray}
\partial_x\Psi=U(x,t,\lambda)\Psi\\
\partial_t\Psi=V(x,t,\lambda)\Psi,
\end{eqnarray}
the so-called zero-curvature condition \cite {Fadd}:
\begin{eqnarray}
\partial_x V - \partial_t U + [V,U] = 0,
\end{eqnarray}
where functions U and V take their values in some Lie algebra $\mathbf{g}$.
The Hamiltonian interpretation \cite{zahfadd} allowed to consider the transformation to the scattering data of the linear problem (1), as a 
transformation to the ``action-angle'' variables in terms of which the problem
is reduced to the system of linear equations. This transform with its inverse give the solution of the Cauchy problem. 
The corresponding family of the integrals of motion can be extracted from the
spectral parameter expansion of the trace of the monodromy matrix of the equation (1). The Poisson brackets of the elements of the monodromy matrix 
$T(\lambda)$ for different values of the spectral parameter often has the following form: 
\begin{eqnarray}
\{T(\lambda)\otimes,T(\mu)\}=[r(\lambda\mu^{-1}),T(\lambda)\otimes T(\mu)],
\end{eqnarray}
where $r$ is a classical $r$- matrix \cite{Fadd}.
Using this relation one can easily show that $\{t(\lambda),t(\mu)\}$
=0, where $t(\lambda)=
trT(\lambda)$, i.e. integrability condition as an existence of the
involutive family of the integrals of motion. 
After the quantization (4) transforms to the RTT-relation:
\begin{eqnarray}
R(\lambda\mu^{-1})(T^{(q)}(\lambda)\otimes I)(I\otimes T^{(q)}(\mu))=
(I\otimes T^{(q)}(\mu))(T^{(q)}(\lambda)\otimes I)R(\lambda\mu^{-1}),
\end{eqnarray}
where $R$ is a quantum R-matrix \cite {leshouches},\cite{sklyan}. 
One can easily show that $[t^{(q)}(\lambda),t^{(q)}(\mu)]=0$ 
as it was on the classical level, i.e. we get the quantum integrability, the 
existence of mutually commuting operators.
The relation (5) is a starting point of the already mentioned second approach
in the theory of the integrable systems.
Using the RTT-relation with the use of different methods, for example 
Algebraic Bethe Ansatz, one can find the spectrum of the transfer-matrix 
$t^{(q)}(\lambda)$ and different correlators.
In order to obtain (5) for the two-dimensional field theory 
systems it is often necessary to consider them on a lattice. 
But for some systems, such as Korteweg-de Vries equation it is possible 
to construct the RTT-relation and find the explicit form of the monodromy matrix using the continuous field theory \cite{1}-\cite{5}.  
The KdV equation also allows quantization in another way with the use of boson-fermion correspondence \cite{pogreb}.\\
\hspace*{5mm}In this paper we extend this class of systems by 
including the supersymmetric generalization of the KdV 
equation \cite {sKdV1}-\cite{sKdV4} and show the 
peculiarities appearing during the quantization in terms of 
continuous field theory of supersymmetric KdV hierarchies. 
The preliminary version of this manuscript was published in \cite{physlett}. 
\subsection{(Super)Conformal Field Theory, its Perturbations
and the Quantum Inverse Scattering Method}
In 1970 it appeared a conjecture \cite {polyakov},
that the field theory corresponding to the fixed (critical) point of the renormalization group has not only scale but conformal invariance \cite {polyakov}.
In two dimensions due to the fact that 2d conformal symmetry is infinite-dimensional and related to the Virasoro algebra: 
\begin{eqnarray}
[L_n,L_m]&=&(n-m)L_{n+m}+\frac{c}{12}(n^3-n)\delta_{n,-m}
\end{eqnarray}
it is possible to classify all the fields in the theory and calculate correlation functions.\\
\hspace*{5mm}Perturbations usually break the conformal symmetry and lead the 
system out of the critical point. However, the perturbations of the 
special type 
called ``integrable'' \cite{int} still preserve the infinite involutive 
algebra of the integrals of motion and thus lead to the integrable theory.\\ 
\hspace*{5mm} In the papers \cite{1}-\cite{5} it was shown that in this case
the problem could be solved with the use of QISM in terms of continuous fields.
It was proposed the following: at first to use conformal symmetry to build basic structures of QISM at the critical point and then with the use of QISM to study the perturbed theory.
 Our object of study is a model, based on a supersymmetric generalization of 
conformal symmetry (superconformal symmetry) \cite{SCFT1}, \cite{SCFT2}. 
 We use its properties to build the quantum monodromy matrix, RTT-relation
and fusion rules for the transfer-matrices in different representations.\\
\hspace*{5mm} As a classical limit of this quantum model it is considered 
the theory of $\widehat{osp}(1|2)$ supersymmetric Korteweg-de Vries equation 
(super-KdV) \cite {sKdV1}-\cite {sKdV4}.
 Two equivalent L-operators, Miura transform and Poisson brackets 
corresponding to the Drinfeld-Sokolov reduction of the affine superalgebra 
$\widehat {osp}(1|2)$ are introduced.
 The associated monodromy matrix is also constructed. Its supertrace is a 
generating function for both local and nonlocal integrals of motion, being in the involution under the Poisson brackets.
 Then with the use of the monodromy matrix the auxiliary matrix 
$\mathbf{L}(\lambda)$ is introduced, the Poisson brackets of which for different values of the spectral parameter have the form (4).\\ 
\hspace*{5mm} After this necessary preparations we move to the quantum theory 
(Sec. 3).
We give the quantum version of Miura transformation, the free field representation of the superconformal algebra \cite{SCFT1}, \cite{SCFT2} and introduce the vertex operators necessary for the construction of the quantum monodromy matrix.\\ 
\hspace*{5mm}
In the quantum case the algebraic structure of the monodromy matrix is described in terms of affine superalgebra $\widehat {osp}_q(1|2)$.
Its representations are constructed in the section 4.
After this we introduce the quantum counterpart of the auxiliary 
$\mathbf{L}$-matrix. It appears that in the quantum case (comparing with classical one) one term is absent in the P-exponent, which is present as in the
monodromy matrix as in auxiliary matrix. Quantum $\mathbf{L}$-matrices satisfy the RTT-relation thus giving the integrability in the quantum case.
Considering the monodromy matrix in different representations of 
$\widehat {osp}_q(1|2)$ we obtain the functional relations (fusion relations) 
for their supertraces, the ``transfer-matrices''. 
In those cases when the deformation parameter is rational (this corresponds to the minimal models of (super)conformal field Theory) fusion relations become 
the closed system of equations, which following the conjecture from 
\cite{1} can be used to find the complete set of the eigenvalues of 
transfer matrices. Moreover we suppose that they could be transformed to the 
Thermodynamic Bethe Ansatz equations \cite{TBA}.\\ 
\section{A Review of Classical Super-KdV Theory} 
In the papers \cite{1}-\cite{5} 
the quantization of the Drinfeld-Sokolov hierarchies of KdV type
related with affine algebras
 $A_1^{(1)}$, $A_2^{(2)}$ и $A_2^{(1)}$.
Our quantum model in the classical limit gives the super-KdV hierarchy 
\cite{sKdV1}-\cite{sKdV4} related with affine superalgebra 
$\widehat {osp}(1|2)$.
  The supermatrix L-operator, corresponding to super-KdV theory
is the following one:
\begin{eqnarray}
\mathcal{L}_F=D_{u,\theta} 
-D_{u,\theta}\Psi h-(iv_{+} \sqrt{\lambda}-\theta\lambda X_{-}),
\end{eqnarray}
where $D_{u,\theta} =\partial_\theta + \theta \partial_u$ 
is a superderivative, variable 
$u$ lies on a cylinder of circumference $2\pi$, $\theta$ 
is  a  Grassmann  variable, $\Psi(u,\theta)=\phi(u) - 
i \theta\xi(u)/\sqrt{2}$ 
is a bosonic superfield; $h,v_+,v_-,X_-,X_+$ are generators of 
$osp(1|2)$ (for more information see \cite{osp1}-\cite{osp2}):
\begin{eqnarray}
[h,X_\pm]&=&\pm 2X_\pm ,\quad [h,v_\pm]=\pm v_\pm ,\quad [X_+,X_-]=h,\\
\lbrack v_\pm,v_\pm\rbrack&=&\pm 2 X_\pm, \quad  [v_{+},v_{-}]=-h, \quad
[X_\pm,v_\mp]= v_\pm, \quad [X_\pm,v_\pm]=0.\nonumber
\end{eqnarray}
Here [,] means supercommutator: $[a,b]=ab-(-1)^{p(a)p(b)}ba$
and the parity $p$ is defined as follows:
$p (v_\pm)=1$, $p (X_\pm)=0$, $p (h)=0$. The 
``fermionic'' operator $\mathcal{L}_F$ considered together 
with a linear problem      
$\mathcal{L}_F\chi(u,\theta)=0$ is equivalent to the ``bosonic'' one:
\begin{eqnarray}
\mathcal{L}_B=\partial_u-\phi'(u)h-\sqrt{\lambda/2}\xi(u)v_{+}-\lambda (X_{+}+X_{-}).
\end{eqnarray}
The fields $\phi$, $\xi$ satisfy the following boundary conditions:
\begin{eqnarray}
\phi(u + 2\pi )&=&\phi(u)+ 2\pi i p,\\
\xi(u + 2\pi )&=&\pm\xi(u), \nonumber
\end{eqnarray}
where ``+'' corresponds to the so-called Ramond (R) sector of the model and   
``--'' to the Neveu-Schwarz (NS) one.
 The Poisson brackets, given by the Drinfeld-Sokolov construction are the
following:
\begin{eqnarray}
\{\xi(u),\xi(v)\}&=&-2\delta(u-v),\\
\{\phi(u),\phi(v)\}&=&\frac{1}{2}\epsilon(u-v).\nonumber
\end{eqnarray}
The L-operators (7), (9) correspond to the super-mKdV, they 
are written in the 
Miura form. Making a gauge
transformation to proceed to the super-KdV L-operator one obtains two fields:
 \begin{eqnarray}\label{eq:Miura}
U(u)&=&-\phi''(u)-\phi'^2(u)-\frac{1}{2}\xi(u)\xi'(u),\\
\alpha(u)&=&\xi'(u) + \xi(u)\phi'(u)\nonumber,
\end{eqnarray}
which generate the superconformal algebra under the 
Poisson brackets:
\begin{eqnarray}
\{U(u),U(v)\}&=&
 \delta'''(u-v)+2U'(u)\delta(u-v)+4U(u)\delta'(u-v),\\
\{U(u),\alpha(v)\}&=&
 3\alpha(u)\delta'(u-v) + \alpha'(u)\delta(u-v),\nonumber\\
\{\alpha(u),\alpha(v)\}&=&
 2\delta''(u-v)+2U(u)\delta(u-v).
\end{eqnarray} 
These brackets describe the second Hamiltonian structure of the super-KdV 
hierarchy. One can obtain evolution equation by taking one of the 
corresponding infinite set of local IM
(they could be obtained by expanding log$(\mathbf{t}_1(\lambda))$, where
$\mathbf{t}_1(\lambda)$ is the supertrace of the monodromy matrix, see below):
\begin{eqnarray}
I^{(cl)}_1&=&\int U(u)\ud u,\\
I^{(cl)}_3&=&\int\Big(\frac{U^2(u)}{2}+\alpha(u)\alpha'(u)\Big)\ud u,
\nonumber\\
I^{(cl)}_5&=&\int\Big((U')^2(u)-2U^3(u)+8\alpha'(u)\alpha''(u)+12
\alpha'(u)\alpha(u)U(u)\Big)\ud u.\nonumber\\
& &   .\qquad.\qquad.\nonumber
\end{eqnarray}
These IM form an involutive set under the Poisson brackets:
\begin{equation}
\{I^{(cl)}_{2k-1},I^{(cl)}_{2l-1}\}=0
\end{equation}
and the $I^{(cl)}_3$ leads to the super-KdV equation
\cite{sKdV1}-\cite{sKdV3}:
\begin{eqnarray}
U_t&=&-U_{uuu}-6UU_u - 6\alpha\alpha_{uu},\\
\alpha_t&=&-4\alpha_{uuu}-6U\alpha_u -3U_u\alpha.\nonumber
\end{eqnarray}
Now let's consider the ``bosonic'' linear problem 
$\pi_s(\mathcal L_B)\chi(u)=0$,
where $\pi_s$ means irreducible representation of $osp(1|2)$ 
labeled by an integer $s\ge 0$ \cite{osp1}-\cite{osp2}. 
We can write the solution of this problem in such a way:
\begin{eqnarray}
\chi(u)&=&\pi_s\Bigg(e^{\phi(u)h}P\exp\int_0^u \ud u'\Big(\sqrt{\lambda/2}\xi(u')v_{+}e^{-\phi(u')}\\
       &+&\lambda (X_{+}e^{-2\phi(u')} + X_{-}e^{2\phi(u')})\Big)\Bigg)\chi_0,
\nonumber
\end{eqnarray}
where $P\exp$ means $P$-ordered exponent and $\chi_0 \in C^{2s+1}$ 
is a constant 
vector.
This could be rewritten in a more general way:
\begin{eqnarray}
\chi(u)&=&\pi_s(\lambda)
\Bigg(e^{-\phi(u)h_{\alpha_0}}P\exp\int_0^u \ud u'\Big(\xi(u')e^{-\phi(u')}e_{\alpha}
+e^{-2\phi(u')}2e_{\alpha}^2 \\
&+& e^{2\phi(u')}e_{\alpha_0}\Big)\Bigg)\chi_0,\nonumber
\end{eqnarray}
where $e_{\alpha}$, $ e_{\alpha_0}$, $h_{\alpha_0}$ 
are part of the Chevalley 
generators of $\widehat {osp}(1|2)$ (see \cite{Tolst}),
which coincide in the evaluation representations $\pi_s(\lambda)$
with $\sqrt{\lambda/2}v_+$, ${\lambda}X_-$, $-h$ correspondingly.  
The associated monodromy matrix then has the form:
\begin{eqnarray}
\mathbf{M}_s(\lambda)&=&\pi_s\Bigg(e^{-2\pi iph_{\alpha_0}}
P\exp\int_0^{2\pi} \ud u'\Big(\xi(u')e^{-\phi(u')}e_{\alpha}
+e^{-2\phi(u')}2e_{\alpha}^2 \\
&+& e^{2\phi(u')}e_{\alpha_0}\Big)\Bigg).\nonumber
\end{eqnarray}
Following \cite{1}-\cite{4} let's introduce auxiliary matrices:
\begin{eqnarray}
\pi_s(\lambda)(\mathbf{L})=\mathbf{L}_s(\lambda)=\pi_s(\lambda)
(e^{\pi ip h_{\alpha_0}})
\mathbf{M}_s(\lambda).
\end{eqnarray}
They satisfy Poisson bracket algebra \cite{Fadd}:
\begin{eqnarray}
\{\mathbf{L}_s(\lambda)\otimes_{,}\mathbf{L}_{s'}(\mu)\}=[\mathbf{r}_{ss'}(\lambda\mu^{-1}),\mathbf{L}_{s}(\lambda)\otimes \mathbf{L}_{s'}(\mu)],
\end{eqnarray}
where $\mathbf{r}_{ss'}(\lambda\mu^{-1})=\pi_s(\lambda)
\otimes\pi_{s'}(\mu)(\mathbf{r})$ 
is the classical trigonometric $\widehat{osp}(1|2)$
r-matrix \cite{Kulish} taken in the corresponding representations:
\begin{eqnarray}
\mathbf{r}(\lambda\mu^{-1})&=&\frac{1}{2}\frac{\lambda\mu^{-1}+
\lambda^{-1}\mu}{\lambda\mu^{-1}-\lambda^{-1}\mu} h\otimes  h+\\
&+&\frac{2}{\lambda\mu^{-1}-\lambda^{-1}\mu}
(X_+\otimes X_- + X_-\otimes X_+)\nonumber\\
&+&\frac{1}{(\lambda\mu^{-1}-\lambda^{-1}\mu)}\Bigg(\sqrt\frac{\mu}{\lambda}
v_+\otimes v_- - \sqrt\frac{\lambda}{\mu}v_-\otimes v_+\Bigg).\nonumber
\end{eqnarray}
From the Poisson brackets for $\mathbf{L}_s(\lambda)$ one obtains that the 
traces
of monodromy matrices $\mathbf{t}_s(\lambda)=str\mathbf{M}_s(\lambda)$ 
commute under the Poisson
bracket:
\begin{eqnarray}\label{eq:Tinv}
\{\mathbf{t}_s(\lambda),\mathbf{t}_{s'}(\mu)\}=0.
\end{eqnarray}
If one expands log($\mathbf{t}_1(\lambda)$) in series of 
$\lambda^{-1}$, one can see 
that the coefficients in this expansion are local IM, as we mentioned earlier.

\section{Free Field Representation of Superconformal Algebra
and Vertex operators}
To quantize the introduced 
classical quantities, we start from a quantum version 
of the Miura transformation (\ref{eq:Miura}),
the so-called free field 
representation of the superconformal algebra \cite{SCFT1}:
\begin{eqnarray}
-\beta^2T(u)&=&:\phi'^2(u):+(1-\beta^2/2)\phi''(u)+\frac{1}{2}:\xi\xi'(u):+\frac{\epsilon\beta^2}{16}\\ 
\frac{i^{1/2}\beta^2}{\sqrt{2}}G(u)&=&\phi '\xi(u)+(1-\beta^2/2)\xi '(u),
\nonumber
\end{eqnarray}
где
\begin{eqnarray}
&&\phi(u)=iQ+iPu+\sum_n\frac{a_{-n}}{n}e^{inu},\qquad
\xi(u)=i^{-1/2}\sum_n\xi_ne^{-inu},\\
&&[Q,P]=\frac{i}{2}\beta^2 ,\quad 
[a_n,a_m]=\frac{\beta^2}{2}n\delta_{n+m,0},\qquad
\{\xi_n,\xi_m\}=\beta^2\delta_{n+m,0}.\nonumber
\end{eqnarray}
Parameter $\beta^2$ plays a role of a semiclassical parameter  
(Planck's constant).
Recall that there are two types of boundary conditions on 
$\xi$: $\xi(u+2\pi)=\pm\xi(u)$. The sign ``+'' corresponds 
to the R sector,the case
when $\xi$ is integer modded, the ``--'' sign corresponds to the NS sector and
$\xi$ is half-integer modded. The variable $\epsilon$ in (20)
is equal to zero
in the R case and equal to 1 in the NS case.\\
One can expand $T(u)$ and $G(u)$ by modes in such a way:
\begin{eqnarray}
T(u)=\sum_nL_{-n}e^{inu}-\frac{\hat{c}}{16},\qquad G(u)=\sum_nG_{-n}e^{inu},
\end{eqnarray}
where the central charge 
$\hat{c}=5-2(\frac{\beta^2}{2}+\frac{2}{\beta^2})$ and $L_n,G_m$ 
generate the superconformal algebra:
\begin{eqnarray}
[L_n,L_m]&=&(n-m)L_{n+m}+\frac{\hat{c}}{8}(n^3-n)\delta_{n,-m}\\
\lbrack L_n,G_m\rbrack&=&(\frac{n}{2}-m)G_{m+n}\nonumber\\
\lbrack G_n,G_m\rbrack&=&2L_{n+m}+\delta_{n,-m}\frac{\hat{c}}{2}(n^2-1/4).
\nonumber
\end{eqnarray}
In the classical limit  $c\to -\infty$ (the same is $\beta^2\to 0$) 
the following substitution:
$
T(u)\to-\frac{\hat{c}}{4}U(u),\quad
G(u)\to-\frac{\hat{c}}{2\sqrt{2i}}\alpha(u),\quad
[,]\to \frac{4\pi}{i\hat{c}}\{,\}
$
reduce the above algebra to the Poisson bracket algebra of 
super-KdV theory.\\ 
\hspace*{5mm}Let $F_p$ be the Fock representation with the vacuum
$|p\rangle$ (highest weight vector). Vector $|p\rangle$ is 
determined by the eigenvalue of $P$ and nilpotency condition of the 
action of the positive modes: 
\begin{eqnarray}
P|p\rangle=p|p\rangle,\quad 
a_n|p\rangle=0, \quad \xi_m|p\rangle=0\quad n,m > 0.
\end{eqnarray}
In the case of the R sector the highest weight becomes doubly degenerate
due to the presence of zero mode $\xi_0$,i.e. there are two ground states
$|p,+\rangle$ and $|p,-\rangle$:
\begin{eqnarray}
|p,+\rangle = \xi_0|p,-\rangle.
\end{eqnarray}
Using the above free field representation of the superconformal algebra
one can obtain that for generic $\hat{c}$ and $p$, $F_p$ is isomorphic to the 
super-Virasoro module with the highest weight vector $|p\rangle$:
\begin{eqnarray}
L_0|p\rangle&=&\Delta_{NS}|p\rangle,\quad \Delta_{NS}=
\Bigg(\frac{p}{\beta}\Bigg)^2 + \frac{\hat{c}-1}{16}
\end{eqnarray}
in the NS sector and module with two highest weight vectors in the Ramond case:
\begin{eqnarray}
L_0|p,\pm\rangle&=&\Delta_{R}|p,\pm\rangle,\quad\Delta_{R}=
\Bigg(\frac{p}{\beta}\Bigg)^2 + \frac{\hat{c}}{16},\\
|p,+\rangle&=&\frac{\beta^2}{\sqrt{2}p}G_0|p,-\rangle.\nonumber
\end{eqnarray}
The space $F_p$, considered as super-Virasoro module, splits in the sum of 
finite-dimensional subspaces, determined by the value of $L_0$:
\begin{eqnarray}
F_p=\oplus^{\infty}_{k=0}F_p^{(k)},\quad
L_0 F_p^{(k)}&=&(\Delta + k) F_p^{(k)}.
\end{eqnarray}
The quantum versions of local integrals of motion should act invariantly on 
the subspaces $F_p^{(k)}$. Thus, the problem of the diagonalization
of IM reduces (in a given subspace $ F_p^{(k)}$) to the finite purely 
algebraic problem,
which however rapidly become very complex for large $k$. It should 
be noted also that in the case of the 
Ramond sector $G_0$ does not commute with IM (even classically), 
so IM mix $|p,+\rangle$ and $|p,-\rangle$.\\
\hspace*{5mm}In the end of this section we introduce another useful notion 
-- vertex operator.
We need two types of them:''bosonic'' and ``fermionic'':
\begin{equation}
V_B^{(a)}=\int \ud\theta\theta :e^{a\Phi}:,\quad V_F^{(b)}=
\int \ud\theta :e^{b\Phi}:,
\end{equation}
where $\Phi(u,\theta)=\phi(u)-\theta\xi(u)$ is a superfield, so
\begin{eqnarray}
V_B^{(a)}=:e^{a\phi}:,\quad V_F^{(b)}=-\frac{ib}{\sqrt{2}}\xi:e^{b\phi}:
\end{eqnarray}
and normal ordering here means that
\begin{eqnarray}
&&:e^{c\phi(u)}:=\\
&&\exp\Big(c\sum_{n=1}^{\infty}\frac{a_{-n}}{n}e^{inu}\Big)
\exp\Big(ci(Q+Pu)\Big)\exp\Big(-c\sum_{n=1}^{\infty}\frac{a_{n}}{n}e^{-inu}
\Big).\nonumber
\end{eqnarray}
\section{Quantum Monodromy Matrix and Fusion Relations}
In this section we will construct the quantum versions of monodromy matrices,
operators $\mathbf{L}_s$ and $\mathbf{t}_s$.\\
   The classical monodromy matrix is based on the $\widehat {osp}(1|2)$ 
affine Lie algebra.
In the quantum case the underlying algebra is quan\-tum 
$\widehat {osp}_q(1|2)$ \cite{Tolst} with $q=e^{i\pi\beta^2}$
and generators, corresponding to even root $\alpha_0$ and odd
root $\alpha$:
\begin{eqnarray} 
&&[h_\gamma,h_{\gamma'}]=0\quad(\gamma,\gamma'=\alpha, d, \alpha_0),\\
&&[e_\beta,e_{\beta'}]=\delta_{\beta,-\beta'}[h_\beta]
\quad(\beta =\alpha,
\alpha_0),\nonumber\\
&&[h_d,e_{\pm\alpha_0}]=\pm e_{\pm\alpha_0},\nonumber\\
&&[h_d,e_{\pm\alpha}]=0,\nonumber\\
&&[h_{\alpha_0},e_{\pm\alpha_0}]=\pm2e_{\pm\alpha_0},\nonumber\\
&&[h_{\alpha_0},e_{\pm\alpha}]=\mp e_{\pm\alpha},\nonumber\\
&&[h_{\alpha},e_{\pm\alpha}]=\pm\frac{1}{2} e_{\pm\alpha},\nonumber\\
&&[h_{\alpha},e_{\pm\alpha_0}]=\mp e_{\pm\alpha_0},\nonumber\\
&&[[e_{\pm\alpha},e_{\pm\alpha_0}]_q,e_{\pm\alpha_0}]_q=0,\nonumber\\
&&[e_{\pm\alpha},[e_{\pm\alpha},[e_{\pm\alpha}[e_{\pm\alpha},[e_{\pm\alpha},
e_{\pm\alpha_0} ]_q]_q]_q]_q]_q=0.\nonumber
\end{eqnarray}
Here $[,]_q$ is the super q-commutator: $[e_a,e_b]=
e_ae_b-q^{(a,b)}(-1)^{p(a)p(b)}e_be_a$ and parity $p$ is defined as 
follows: $p(h_{\alpha_0})=0,\quad 
p(h_{\alpha})=0,\quad p(e_{\pm\alpha_0})=0 ,
\quad p(e_{\pm\alpha})=1$.
Also, as usual, $[h_{\beta}]=\frac{q^{h_\beta}-q^{-h_\beta}}{q-q^{-1}}$.
The finite dimensional representations $\pi_s^{(q)}(\lambda)$ of 
$\widehat {osp}_q(1|2)$ 
can be characterized by integer number s and have the following explicit 
form:
\begin{eqnarray}
h_{\alpha_0}|j,m\rangle&=&2m|j,m\rangle,\\
e_{\alpha_0}|j,m\rangle&=&\lambda\sqrt{[j-m][j+m+1]}|j,m+1\rangle,\nonumber\\
e_{-\alpha_0}|j,m\rangle&=&\lambda^{-1}
\sqrt{[j+m][j-m+1]}|j,m-1\rangle,\nonumber\\
e_\alpha|j,m\rangle&=&\sqrt{\lambda}
((-1)^{-2j}\sqrt{\alpha(j)[j-m+1]}|j+1/2,m-1/2\rangle\nonumber\\
&+&\sqrt{\alpha(j-1/2)[j+m]}|j-1/2,m-1/2\rangle),\nonumber\\
e_{-\alpha}|j,m\rangle&=&\sqrt{\lambda}^{-1}
(-\sqrt{\alpha(j)[j+m+1]}|j+1/2,m+1/2\rangle\nonumber\\
&-&(-1)^{2j}\sqrt{\alpha(j-1/2)[j-m]}|j-1/2,m+1/2\rangle),\nonumber\\
h_{\alpha_0}&=&-2h_{\alpha},\nonumber\\
 h_d&=&\frac{1}{2}\lambda\frac{d}{d\lambda}+
\frac{1}{4}h_{\alpha_0},\nonumber
\end{eqnarray}
where $j=0, 1/2, ..., s/2$, $m= -j, -j+1 ..., j$.\\
The normalization coefficients
\begin{eqnarray}
\alpha(j)&=&\frac{[j+1][j+1/2][1/4]}{[2j+2][2j+1][1/2]}
\Bigg((-1)^{s-2j+1}
\frac{[s+3/2]}{[s/2+3/4]}+\frac{[j+3/2]}{[j/2+3/4]}\bigg)\nonumber\\
\end{eqnarray}
are defined by the recurrence relation:
\begin{eqnarray}
\alpha(j)\frac{[2j+2]}{[j+1]} + \alpha(j-1/2)\frac{[2j]}{[j]}=1,\quad
\alpha(s/2)=0 .
\end{eqnarray}
It is not hard to see that in the classical limit $q\to 1$ $\alpha(s/2-k)=0$,
if $k < s/2$ is a nonnegative 
integer and 
$\alpha(s/2-k)=1/2$, if $k < s/2$ is nonnegative half-integer. Using this fact 
one can obtain 
that this representation in the classical limit appears to be a direct sum
of finite dimensional irreducible representations of $\widehat {osp}(1|2)$:
\begin{eqnarray} 
\pi_s^{(1)}(\lambda)=\oplus_{k=0}^{[s/2]}\pi_{s-2k}(\lambda).
\end{eqnarray}
In this sum k runs through integer numbers.
One can  notice that the structure of irreducible finite dimensional 
representations of $\widehat {osp}_q(1|2)$ 
is similar to those of $(A_2^{(2)})_q$ \cite{4}. This is the consequence of 
the coincidence of their Cartan matrices.\\
\hspace*{5mm}After these preparations we are ready to introduce the quantum 
counterparts of
$\mathbf{L}_s$ operators:
\begin{eqnarray}
\mathbf{L}_s^{(q)}&=&\pi_s^{(q)}(\lambda)(\mathbf{L}^{(q)})\\
&=&\pi_s^{(q)}\Bigg(e^{-i\pi Ph_{\alpha_0}}P\exp\bigg(\int_0^{2\pi} \ud u
\Big(:e^{2\phi(u)}:e_{\alpha_0}
+\xi(u):e^{-\phi(u)}:e_{\alpha}\Big)\bigg)\Bigg).\nonumber
\end{eqnarray}
 One can see that one term is missing in the P-exponent in comparison
with the classical case (20), (21). Disappearing of terms corresponding to 
composite roots takes place also for general supersymmetric KdV hierarchies  
(we will return to this elsewhere).\\
\hspace*{5mm}Analyzing the singularity properties of the integrands in P-exponent
of $\mathbf{L}^{(q)}_s(\lambda)$ one can find that the integrals are 
convergent for values of $\hat c$ from the interval:
\begin{eqnarray}
-\infty<\hat c< 0
\end{eqnarray}
With the use of regularization P-exponent can be continuated on a wider region
of $\hat c$.\\
\hspace*{5mm}
Now let's prove that in the classical limit $\mathbf{L}^{(q)}$ will 
coincide with $\mathbf{L}$.\\
 First let's analyse the products of the operators we have in the P-exponent.
The product of the two fermion operators can be written in such a way:
\begin{eqnarray}
\xi(u)\xi(u')=:\xi(u)\xi(u'):-i\beta^2 
\frac{e^{-\kappa \frac{i}{2}(u-u')}}
{e^{\frac{i}{2}(u-u')}-e^{-\frac{i}{2}(u-u')}}.
\end{eqnarray}
where $\kappa$ is equal to zero in the NS sector and equal to 1 in 
the R sector.
for vertex operators the corresponding operator product is:
\begin{eqnarray}
:e^{a\phi(u)}::e^{b\phi(u')}:=
(e^{\frac{i}{2}(u-u')}-e^{-\frac{i}{2}(u-u')})^{\frac{ab\beta^2}{2}}
:e^{a\phi(u)+b\phi(u')}:,
\end{eqnarray}
where 
\begin{eqnarray}
:e^{a\phi(u)+b\phi(u')}:=&&
\exp\Big(a\sum_{n=1}^{\infty}\frac{a_{-n}}{n}e^{inu}+
b\sum_{n=1}^{\infty}\frac{a_{-n}}{n}e^{inu'}\Big)\\
&& \exp\Big(ai(Q+Pu)+bi(Q+Pu')\Big)\nonumber\\
&& \exp\Big(-a\sum_{n=1}^{\infty}\frac{a_{n}}{n}e^{-inu}-b\sum_{n=1}^{\infty}\frac{a_{n}}{n}e^{-inu'}\Big).\nonumber
\end{eqnarray}
It would be more useful to rewrite these products picking out the singular part:
\begin{eqnarray}
&&\xi(u)\xi(u')=
-\frac{i\beta^2}{(iu-iu')}+\sum_{k=1}^{\infty}c_k(u)(iu-iu')^k,\\
&&:e^{a\phi(u)}::e^{b\phi(u')}:=(iu-iu')^{\frac{ab\beta^2}{2}}
(:e^{(a+b)\phi(u)}:+\sum_{k=1}^{\infty}d_k(u)(iu-iu')^k),
\end{eqnarray}
where $c_k(u)$ and $d_k(u)$ are operator-valued functions of $u$.
Now let's return to the  $\mathbf{L}^{(q)}(\lambda)$ operator, which
one can express it in the following way:
\begin{eqnarray}
\mathbf{L}^{(q)}&=&
e^{-i\pi Ph_{\alpha_0}}\lim_{N\to\infty}\prod_{m=1}^{N}\tau_m^{(q)},\\
\tau_m^{(q)}&=&P\exp\int_{x_{m-1}}^{x_{m}}\ud u K(u),\nonumber\\
K(u)&\equiv&:e^{2\phi(u)}:e_{\alpha_0}+
\xi(u):e^{-\phi(u)}:e_{\alpha}. \nonumber
\end{eqnarray}
Here we have divided  the interval $[0,2\pi]$ into small 
intervals $[x_m,x_{m+1}]$
with $x_{m+1}-x_m=\Delta=2\pi/N$. 
Let's look on the behaviour of the first two iterations when $\beta^2\to 0$:
\begin{eqnarray}
\tau_m^{(q)}=1+\int_{x_{m-1}}^{x_{m}}\ud u K(u) +
\int_{x_{m-1}}^{x_{m}}\ud u K(u)\int_{x_{m-1}}^{u}\ud u'K(u')+
O(\Delta^2). 
\end{eqnarray}
It appears, that in  $\beta^2\to 0$ limit terms from the second iteration
can give contribution to the first one. To see this let's  consider the 
expression that comes from the second iteration:
\begin{eqnarray}
-\int_{x_{m-1}}^{x_{m}}\ud u\xi(u)\int_{x_{m-1}}^{u}\ud u'\xi(u'):e^{-\phi(u)}:
:e^{-\phi(u')}: e_{\alpha}^2.
\end{eqnarray}
Using the above operator products and seeking the terms of order
$\Delta^{1+\beta^2}$ (only those can give us the first iteration terms in 
$\beta^2\to 0$ limit) one obtains that their contribution is:
\begin{eqnarray}
&&i\beta^{2}\int_{x_{m-1}}^{x_{m}}\ud u\int_{x_{m-1}}^{u}\ud u'
(iu-iu')^{\frac{\beta^2}{2}-1}:e^{-2\phi(u)}: e_{\alpha}^2\\
&=&2\int_{x_{m-1}}^{x_{m}}\ud u:e^{-2\phi(u)}:(iu-ix_{m-1})^{\frac{\beta^2}{2}}
e_{\alpha}^2.\nonumber
\end{eqnarray}
Considering this in the classical limit we recognize the familiar terms from 
$\mathbf{L}$:
\begin{eqnarray}
\tau_m^{(1)}=1+\int_{x_{m-1}}^{x_{m}}\ud u\Big(\xi(u)e^{-\phi(u)}e_{\alpha}
+ e^{2\phi(u)}e_{\alpha_0} +
e^{-2\phi(u)} 2 e_{\alpha}^2\Big)+O(\Delta^2).
\end{eqnarray}Collecting all
$\tau_m^{(1)}$ one obtains the desired result:
\begin{eqnarray}
\mathbf{L}^{(1)}=\mathbf{L}.
\end{eqnarray}
Recalling the structure of 
$\widehat{osp}_q(1|2)$ representations we get:
\begin{eqnarray}
\mathbf{L}_s^{(1)}(\lambda)=\sum_{k=0}^{[s/2]}\mathbf{L}_{s-2k}(\lambda).
\end{eqnarray}
\hspace*{5mm}Using the properties of quantum R-matrix \cite{leshouches} it  
follows
that $\mathbf{R}\Delta(\mathbf{L}^{(q)})=\Delta^{op}(\mathbf{L}^{(q)})
\mathbf{R}$, where $\Delta$ and $\Delta^{op}$ are coproduct and 
opposite coproduct of $\widehat{osp}_q(1|2)$ \cite{Tolst} correspondingly.
Factorizing $\Delta(\mathbf{L}^{(q)})$ and 
$\Delta^{op}(\mathbf{L}^{(q)})$ ,  
according to the properties of vertex operators and P-exponent, 
we get the so called RTT-relation 
\cite{leshouches},\cite{sklyan}:
\begin{eqnarray}
&&\mathbf{R}_{ss'}(\lambda\mu^{-1})
\Big(\mathbf{L}_s^{(q)}(\lambda)\otimes \mathbf{I}\Big)\Big(\mathbf{I}
\otimes \mathbf{L}_{s'}^{(q)}(\mu)\Big)\\
&&=(\mathbf{I}\otimes \mathbf{L}_{s'}^{(q)}(\mu)\Big)
\Big(\mathbf{L}_s^{(q)}(\lambda)\otimes \mathbf{I}\Big)\mathbf{R}_{ss'}
(\lambda\mu^{-1}), \nonumber
\end{eqnarray}
where $\mathbf{R}_{ss'}$ is the trigonometric solution of the corresponding 
Yang-Baxter equation \cite{Kulish} which acts in the space $\pi_s(\lambda)
\otimes\pi_{s'}(\mu)$.
Let's define now the ``transfer matrices'' which are the quantum counterparts of
the traces of monodromy matrices: 
\begin{eqnarray} 
\mathbf{t}_s^{(q)}(\lambda)=str\pi_s(\lambda) (e^{-i\pi ph_{\alpha_0}}
\mathbf{L}_s^{(q)}) .
\end{eqnarray}
According to the RTT-relation we obtain:
\begin{eqnarray}\label{eq:qTinv}
[\mathbf{t}_s^{(q)}(\lambda),\mathbf{t}_{s'}^{(q)}(\mu)]=0.
\end{eqnarray}
For the first nontrivial representation (s=1) it is quite easy 
to find the expression for $\mathbf{t}_1^{(q)}(\lambda)\equiv
\mathbf{t}^{(q)}(\lambda)$:
\begin{eqnarray}
\mathbf{t}^{(q)}(\lambda)=
1-2\cos(2\pi iP) + \sum^{\infty}_{n=1}\lambda^{2n}Q_n,
\end{eqnarray}where 
$Q_n$ are nonlocal integrals of motion. 
Following (\ref{eq:qTinv}), $\mathbf{t}^{(q)}(\lambda)$ is
the generation function for mutually
commuting nonlocal conservation laws:
\begin{eqnarray}
[Q_n,Q_m]=0.
\end{eqnarray}
We also suppose that $\mathbf{t}^{(q)}(\lambda)$ 
generates local IM in the 
classical case.
Using (\ref{eq:qTinv}) again one obtains, expanding
log($\mathbf{t}^{(q)}(\lambda)$):
\begin{eqnarray} \label{eq:IMinv}
[Q_n,I^{(q)}_{2k-1}]=0,\qquad
[I^{(q)}_{2l-1},I^{(q)}_{2k-1}]=0.
\end{eqnarray}
First few orders of expansion in $\lambda^2$ of 
$\mathbf{t}_s^{(q)}(\lambda)$ operators the following fusion
relation can be obtained:
\begin{eqnarray}\label{eq:fusion}
\mathbf{t}^{(q)}_s(q^{1/4}\lambda)\mathbf{t}^{(q)}_s(q^{-1/4}\lambda)=
\mathbf{t}^{(q)}_{s+1}(q^{\frac{1}{2\beta^2}}\lambda)
\mathbf{t}^{(q)}_{s-1}(q^{\frac{1}{2\beta^2}}\lambda)+
\mathbf{t}^{(q)}_s(\lambda).
\end{eqnarray}This result reminds the fusion relation for 
$(A_2^{(2)})_q$ case \cite{4}. Such correspondence
should not seem to be extraordinary because of the coincidence 
of their Cartan matrices and similarities in the spectra of representations 
as we mentioned above.\\

We are grateful to P.I. Etingof, M.A. Semenov-Tian-Shansky, 
F.A. Smirnov and V.O. Tarasov for useful discussions.
The work was supported by the Dynasty Foundation (AMZ) and 
RFBR grant 03-01-00593 (PPK).

\end{document}